\documentclass[superscriptaddress,reprint,citeautoscript,prb]{revtex4-1}
\usepackage{graphicx} 
\usepackage[margin=0.9in]{geometry} 
\usepackage{amsmath}
\usepackage{hyperref} 
\usepackage{multirow}
\usepackage{newtxtext}                                        
\usepackage{booktabs}
\usepackage{xcolor} 
\usepackage{hyperref}
\usepackage{mathtools}
\usepackage[normalem]{ulem}
\hypersetup{colorlinks, 
	linkcolor={blue!75!black!80!yellow},
	citecolor={blue!75!black!80!yellow}, 
	urlcolor={blue!75!black!80!yellow}
	}
\usepackage[cmintegrals]{newtxmath}
\makeatletter \renewcommand\@make@capt@title[2]{\@ifx@empty\float@link{\@firstofone}{\expandafter\href\expandafter{\float@link}}\sffamily{\textbf{#1}}\@caption@fignum@sep#2 }

 \newcommand{\HarvardSEAS}{John A. Paulson School of Engineering and Applied
Sciences, Harvard University, Cambridge, MA, USA}
\newcommand{\HarvardCCB}{Department of Chemistry and Chemical Biology, Harvard
University, Cambridge, MA, USA} 
\newcommand{\MIT}{Department of Electrical Engineering and Computer Science,
Massachusetts Institute of Technology, Cambridge, MA, USA}
\newcommand{\ccq}{Center for Computational Quantum Physics, Flatiron Institute, New York, NY, USA}

\usepackage[textwidth=1.5cm,textsize=footnotesize]{todonotes}

\newcommand{\chh}{C\textsubscript{2h}}
\newcommand{\dddd}{D\textsubscript{3d}}
\newcommand{\Vm}{V\textsuperscript{--}}
\newcommand{\Vn}{V\textsuperscript{0}}

\begin{document} 

\author{Christopher J. Ciccarino}
\thanks{These authors contributed equally}
\affiliation{\HarvardSEAS}\affiliation{\HarvardCCB}
\author{Johannes Flick}
\thanks{These authors contributed equally}
\affiliation{\HarvardSEAS}
\affiliation{\ccq}
\author{Isaac B. Harris}
\affiliation{\HarvardSEAS}\affiliation{\MIT}
\author{Matthew E. Trusheim}
\affiliation{\HarvardSEAS}
\author{Dirk R. Englund}\affiliation{\MIT}
\author{Prineha Narang}\email{prineha@seas.harvard.edu}\affiliation{\HarvardSEAS}

\title{Strong Spin-Orbit Quenching \emph{via} the 
Product Jahn-Teller Effect\\ 
in Neutral Group IV Artificial Atom Qubits in Diamond}

\date{\today}

\begin{abstract} 
Artificial atom qubits in diamond have emerged as leading candidates for a range of solid-state quantum systems, from quantum sensors to repeater nodes in  memory-enhanced quantum communication.
Inversion-symmetric group IV vacancy centers, comprised of Si, Ge, Sn and Pb dopants, hold particular promise as their neutrally charged electronic configuration results in a ground-state spin triplet, enabling
long spin coherence above cryogenic temperatures.
However, despite the tremendous interest in these defects, a theoretical understanding of the electronic and spin structure of these centers remains elusive. 
In this context, we predict the ground- and excited-state properties 
of the neutral group IV color centers from
first principles. 
We capture the product Jahn-Teller effect found in the excited state manifold to second order in electron-phonon coupling,
and present a non-perturbative treatment of the effect of spin-orbit coupling. Importantly, we find that spin-orbit splitting
is strongly quenched due to the dominant Jahn-Teller
effect, with the lowest optically-active
$^3E_u$ state weakly split into $m_s$-resolved states.
The predicted complex vibronic
spectra of the neutral group IV color centers are essential for their experimental identification and have key implications for use of these systems in quantum information science. 
\end{abstract}

\maketitle

Artificial atoms in diamond are promising candidates for 
a wide variety of quantum 
technologies~\cite{atature_material_2018,awschalom_quantum_2018,aharonovich_solid-state_2016, rogers_multiple_2014, Weber2010},
including as quantum repeaters for long-range quantum
networks~\cite{kalb_entanglement_2017,humphreys_deterministic_2018}.
Many milestones have been reached using the nitrogen-vacancy (N\Vm)~
center~\cite{doherty_nitrogen-vacancy_2013,rozpedek_near-term_2019}
and more recently the Si\Vm~\cite{Rogers:2014,Hepp2014a,lemonde2018,evans2018,Sukachev2017}.
Further exploration of novel 
defect candidates has included 
the Ge\Vm~\cite{palyanov:GeV:2015,siyushev:GeV:2017,fan:GeV:2018,Bhaskar2017},
Sn\Vm~\cite{iwasaki:SnV:2017,trusheim2020,rugar2019,gorlitz2019}, Pb\Vm~\cite{trusheim_lead-related_2019,tchernij:PbV:2018}
and Si\Vn~\cite{green:SiV0:2019,green:siv0:2017,rose:SiV0:2018}, all of which 
have been observed experimentally and described theoretically~\cite{thiering:gIV:2018,thiering:gIV0:2019}.
The neutrally-charged Si\Vn\ has symmetry analogous
to the Si\Vm, but its missing electron gives
rise to a triplet ground state as found in the N\Vm,
with the corresponding potential for both long spin coherence times and symmetry-protected optical transitions. 
Theoretical work
has postulated the remaining group IV neutral (I\Vn) centers~\cite{thiering:gIV0:2019} (Ge\Vn, Sn\Vn, Pb\Vn)
and described the negatively-charged group III defect centers~\cite{harris2019} as isoelectronic to the Si\Vn. Calculations suggest 
that all of these defect candidates are thermodynamically more likely to exist in intrinsic diamond
than the Si\Vn, which requires p-type doping~\cite{rose:SiV0:2018}. 
Within this growing space of candidate artificial atom qubits, an \emph{ab initio} understanding
of the level structure is required to harness the advantages of
each emitter in quantum science~\cite{advFuncMat}.

Accurate descriptions of artificial atoms
in diamond can be particularly challenging 
because of the dominant Jahn-Teller (JT) distortions~\cite{bersuker1990} present.
In such systems, the total energy of
a JT-unstable electronic configuration
is lowered as a result of the coupling of
the electronic structure to nuclear motion, 
introducing
electron-phonon interactions.
In the case of group I\Vn\ defects,
the excited state exhibits a product
Jahn-Teller (pJT) effect which
results from simultaneous 
Jahn-Teller instabilities in
two orbitals~\cite{thiering:gIV0:2019,qiu:2007,qiu_product_2001,bersuker2017}. 
The pJT interaction leads 
to either a dynamical or 
static JT effect, or a mixture of both.
In the case of a dynamical JT distortion,
the system is best described as a collective
electron-vibration (vibronic) system. 
This strong coupling of electronic and vibrational
states may modify electronic 
observables, for example a quenching of spin-orbit (SO) coupling (SOC).

Including the pJT effect is therefore critical for predictions of
the zero-phonon line (ZPL) transition energies and the 
excited-state level structure. 
Previous work has found that describing pJT 
interactions to first order in coupling
explains the observed energy splitting~\cite{green:SiV0:2019} between
the optically-bright $E_u$ and dark $A_{2u}$
states for Si\Vn~\cite{thiering:gIV0:2019}.
An important effect to consider,
particularly for the heavier group I\Vn\ defects,
is the role of spin-orbit interactions,
as these defects can have coupling constants
on the order of 100s of meV~\cite{thiering:gIV:2018}. 
The interplay of SOC interactions
and JT physics in the excited-state of group
I\Vn\ centers has significant impact on the expected SO behavior
if the JT effect couples the electrons
and phonons strongly, as we find.

In this \emph{Letter}, we describe the combined impact of spin-orbit and 
Jahn-Teller interactions in the neutral group IV 
centers in diamond from first principles. We describe the product Jahn-Teller
effect to second order in electron-phonon coupling and find a large second order energy shift. 
Importantly, the effects of spin-orbit coupling are included
non-perturbatively and splittings
are found to be an order of magnitude 
smaller than expected for a purely electronic system
as a result of the  
JT interaction. 
These fine structure details reveal new physics of color center qubits
in diamond and present a pathway to identify Ge\Vn, 
Sn\Vn\ and Pb\Vn\ experimentally. 

\begin{figure}
\includegraphics[width=\columnwidth]{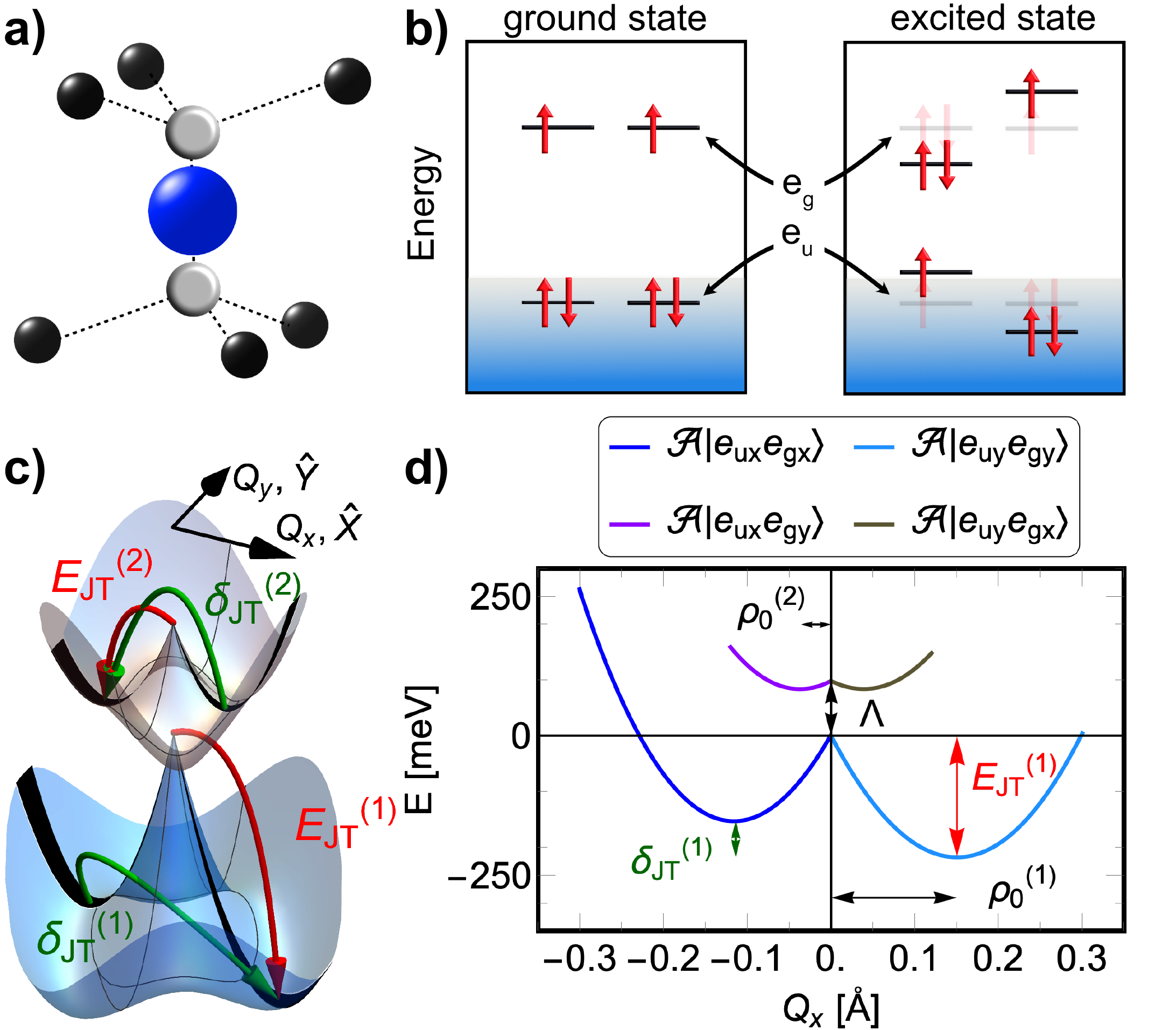}
\caption{ 
\textbf{(a)} Lattice configuration
of the group I\Vn\ defects,
in which the impurity
atom (blue) sits between two vacant
carbon sites (grey).
\textbf{(b)} Simplified
energy level diagram showing the energy 
location of the doubly-degenerate
$e_u$ and $e_g$ orbitals relative to the 
band gap of bulk diamond. The ground
state is a spin triplet and the 
corresponding excited state undergoes a 
symmetry-breaking pJT 
distortion (right) as a result of orbital
instabilities in both 
the $e_u$ and $e_g$ orbitals.
\textbf{(c)} Potential energy surfaces
computed for the pJT
system including effects up to 2\textsuperscript{nd} order
in coupling. 
Here we label the energy instability
by $E_{\mathrm{JT}}^{(i)}$ for 
the result of constructive ($i=1$)
and destructive ($i=2$) interference of the 
two orbital branches. The axial asymmetry
arises from a second order effect denoted
similarly by the parameter $\delta_{\mathrm{JT}}^{(i)}$.
The black curves indicate 1D cuts through the 2D
$(Q_x,Q_y)$ distortion space which allows us to fully
parameterize the system. \textbf{(d)}
DFT-obtained potential energy surfaces
along these 1D cuts for the Sn\Vn\ defect. 
The \dddd\ high-symmetry point ($Q_x = 0$ \AA) is found to be unstable in two surfaces, consistent
with the pJT picture.
We also label 
the displacement amplitudes $\rho_0^{(i)}$ from
the \dddd\ to the \chh\ minima. The splitting
$\Lambda$ is a result of static electronic correlation.
All values are tabulated in Table~\ref{tab:pjt}.
}
\label{fig:fig1}
\end{figure}

The group IV centers in diamond adopt a split-vacancy configuration
within the diamond lattice where the dopant group IV atom
sits between two vacant carbon sites,
as shown in Fig.~\ref{fig:fig1}(a)
and denoted by the point group \dddd.
The defect introduces localized electronic orbitals comprised of 
the dangling 
bonds of the nearby carbon atoms, which can 
be captured using density 
functional theory (DFT)~\cite{vasp} (see SI for computational
details) and are labeled by their symmetry.
The energetically-relevant orbitals
are of 
$e_u$ and $e_g$ character and 
exist near and above the valence band of intrinsic 
diamond, respectively, shown schematically
in Fig.~\ref{fig:fig1}(b). Both the
$e_u$ and $e_g$ orbitals are doubly-degenerate 
and can be further labeled by their spatial orientation,
i.e., $\{e_u\}=\{e_{ux},e_{uy}\}$
and similarly $\{e_g\}=\{e_{gx},e_{gy}\}$.
Including spin, these levels combined can host up to 
eight electrons. 
For 
group I\Vn\ centers, six electrons are present in the $(e_ue_g)$
manifold. Equivalently, we can describe these electronic
states in the basis of two defect-bound holes. We choose
to adopt this convention for the remainder of this \emph{Letter}.

The ground state has the hole configuration
$e_g^2$ ($e_{gx}^1e_{gy}^1$),
and prefers the triplet $S=1$
spin configuration. 
The total defect
wavefunction is of $^3A_{2g}$
symmetry, and is directly obtained from
electronic structure calculations. In constructing 
the total wavefunction, given the symmetric
triplet spin component, we ensure that
the orbital wavefunction is antisymmetrized;
this is given by the $\mathcal{A}$ symbol.
The ground state orbital wavefunction
can be written as
$\mathcal{A}|e_{gx}e_{gy}\rangle = 
1/\sqrt{2}\left(  e_{gx} (\mathbf{r}_1)
 e_{gy} (\mathbf{r}_2) - e_{gx} (\mathbf{r}_2)
 e_{gy}  (\mathbf{r}_1)\right)$. 
In the excited electronic configuration,
one hole moves from an $e_g$ to an
$e_u$ orbital. 
Unlike in the ground state, 
there exist four distinct hole occupations
with this $e_g^1e_u^1$ configuration.
The antisymmetrized orbital wavefunctions are given by
$    \mathcal{A} |e_{ux}e_{gx} \rangle$,
    $\mathcal{A} |e_{uy}e_{gx} \rangle$,
    $\mathcal{A} |e_{ux}e_{gy} \rangle$,
    and
    $\mathcal{A} |e_{uy}e_{gy} \rangle$.
We can construct the irreducible representations
of the triplet subspace as linear combinations of these orbital
states, as has been done previously~\cite{thiering:gIV0:2019}. 

Each of these antisymmetrized states obtained from our \emph{ab initio}
calculations are Jahn-Teller unstable, in that they energetically prefer a configuration with
the lower symmetry \chh\ point group to that with the higher symmetry \dddd\
point group.
The nuclear motion associated with these distortions 
is a result of interactions with phonon modes of symmetry
$E_g$. 
In contrast with the single JT system ($E_g \otimes e$), 
the JT distortion found in the excited state
of group I\Vn\ systems is due to simultaneous JT interactions
in both the $e_u$ and $e_g$ orbitals. This collective product Jahn-Teller
behavior is denoted by $E_g \otimes e_u \otimes e_g$
and shown schematically in the right panel of 
Fig.~\ref{fig:fig1}(b).
Previous work has covered the single JT to second order as well as the 
pJT~\cite{bersuker1990,bersuker2017,qiu:2007}
to first order in electron-phonon coupling. Here, we describe the coupling
of the two electronic states with the
$E_g$-type vibrational mode to second order
in vibrational coupling. The Hamiltonian
for this interaction can be written
as:
\begin{align}
    \hat{\mathrm{H}}_{\mathrm{pJT}}^{(2)} 
    = 
    & F_u \left( \hat{X}\hat{\sigma_z}\otimes \hat{\sigma}_0
   - \hat{Y} \hat{\sigma}_x \otimes \hat{\sigma}_0
   \right)  \nonumber \\
   & + F_g \left( \hat{X}\hat{\sigma_0}\otimes \hat{\sigma}_z 
   - \hat{Y} \hat{\sigma}_0 \otimes \hat{\sigma}_x
   \right) \nonumber\\
   & + G_u
   \left( \left(\hat{X}^2 - \hat{Y}^2 \right) 
     \hat{\sigma}_z \otimes \hat{\sigma}_0   
    + 2
    \hat{X} \hat{Y}  \hat{\sigma}_x \otimes 
    \hat{\sigma}_0 
    \right) 
    \nonumber\\
    & + 
    G_g 
    \left(
    \left( \hat{X}^2 - \hat{Y}^2 \right) 
    \hat{\sigma}_0 \otimes \hat{\sigma}_z
    + 2 \hat{X} \hat{Y} \hat{\sigma}_0 \otimes 
    \hat{\sigma}_x
    \right).
\label{eq:pjt}
\end{align}
The first two lines represent linear
coupling with coupling constants $F_{u/g}$ while the latter two represent
quadratic coupling terms with coupling constants $G_{u/g}$ for
both the $e_g$ and $e_u$ orbital branches.
The nuclear component of the Hamiltonian is written
with $\hat{X}$ and $\hat{Y}$ representing
bosonic operators for the phonons given by
$\{\hat{X},\hat{Y}\} = 
(\hat a_{\{x,y\}}^\dag + \hat a_{\{x,y\}})/\sqrt{2}$
and the electronic component in terms of $\hat{\sigma}_{i}$
which are the standard Pauli and unit matrices acting on the $e_u\otimes e_g$ subspace.
The Hamiltonian in Eq.~\ref{eq:pjt} is defined within the
single-excitation 2-particle hole manifold,
therefore the basis states are
    $\mathcal{A} |e_{ux}e_{gx} \rangle$,
    $\mathcal{A} |e_{uy}e_{gx} \rangle$,
    $\mathcal{A} |e_{ux}e_{gy} \rangle$,
    and
    $\mathcal{A} |e_{uy}e_{gy} \rangle$,
which are captured from electronic structure calculations.

In the pJT case, two independent solutions 
which are unstable at the high-symmetry point can exist. 
One corresponds to the
constructive interference
of the two JT distortions 
$(\sim (F_g + F_u)^2)$
and the other to the destructive
interference 
$(\sim (F_g - F_u)^2)$,
as shown in Fig.~\ref{fig:fig1}(c).
To find the coupling
constants and solve for the 
coupled vibronic states, we obtain
displacement 
$\rho^{(i)}_{0}$ and energy
$E_{\mathrm{JT}}^{(i)}$, 
$\delta_{\mathrm{JT}}^{(i)}$ parameters
from the defect potential energy surfaces (PES)
computed from first principles electronic structure,
where $i=1,2$ for the constructive
and destructive pJT, respectively.
For the Sn\Vn\
color center we show the resulting adiabatic 
PES as a one-dimensional
cut along $Q_y=0$ in Fig.~\ref{fig:fig1}(d).
In principle the PES are 
two-dimensional, with the minima being threefold
degenerate (see Fig.~\ref{fig:fig1}(c)).
However, due to the symmetry of the PES, this 1D cut completely
parameterizes the pJT Hamiltonian.
For additional details
on connecting the coupling 
constants in Eq.~\ref{eq:pjt}
to our calculations, 
refer to the SI.

In these defect systems electronic 
correlation $\hat{\mathrm{W}}$ plays a role in splitting the electronic states
for reasons distinct from the Jahn-Teller physics. 
This correlation can be incorporated
along the lines of previous work~\cite{thiering:gIV0:2019},
leading to the following total Hamiltonian for the system:
\begin{equation}
\hat{\mathrm{H}} = \hat{\mathrm{H}}_{\mathrm{osc}} + \hat{\mathrm{H}}_{\mathrm{pJT}}^{(2)} + \hat{\mathrm{W}}.
\label{eq:pjtandcorrelation}
\end{equation}
Here, $\hat{\mathrm{H}}_{\mathrm{osc}}=\hbar \omega_E \sum_{i=x,y} \left(\hat{a}^\dagger_i \hat{a}_i + 1/2 \right)$ is the two-dimensional
harmonic oscillator Hamiltonian for the $E_g$ phonon
modes of energy $\hbar \omega_E$.

\begin{table}
\begin{tabular}{lcccc}
\hline
 &  Si\Vn & Ge\Vn & Sn\Vn & Pb\Vn \\
\hline
$\rho_0^{(1)}$ [\AA] & 0.171 & 0.166 & 0.154 & 0.145 \\
$\rho_0^{(2)}$ [\AA] & -0.006 & -0.022 & -0.038 & -0.051 \\
$\hbar \omega_E$  [meV] & 87.3 & 86.6 & 87.7 & 90.8 \\
$\Lambda$ [meV] & 81.6 & 86.4 & 98.2 & 112.5 \\
$E_{\mathrm{JT}}^{(1)}$  [meV] & 258 & 244 & 217 & 200 \\
$\delta_{\mathrm{JT}}^{(1)}$  [meV] & 82.2 & 75.5 & 63.5 & 64.5 \\
$E_{\mathrm{JT}}^{(2)}$ [meV] & 0.289 & 4.61 & 14.9 & 29.9 \\
$\delta_{\mathrm{JT}}^{(2)}$  [meV] & 0.147 & 0.307 & 0.226 & 2.18 \\
$\gamma^{(1)}$ [meV] & 7.18 & 7.59 & 8.96 & 10.4 \\
$\gamma^{(2)}$ [meV] & 3.21 & 4.06 & 6.22 & 7.90 \\
ZPL ($^3E_u$) [eV] & 1.361 & 1.813 & 1.833 & 2.216 \\
\hline 
$\gamma^{(2)}$ + SOC [meV] & 3.17 & 3.77 & 4.76 & 2.03 \\
ZPL ($^3E_u$) + SOC [eV] & 1.361 & 1.812 & 1.825 & 2.170 \\
$p_u$ & 0.012 & 0.017 & 0.032 & 0.043 \\
$p_g$ & 0.012 & 0.012 & 0.023 & 0.040 \\
$\lambda_u + \lambda_g$ [meV] & 0.089 & 0.622 & 3.15 & 11.31 \\
\hline
\end{tabular}
\caption{
We determine the parameters $\rho_0^{(i)}$,
$E_{\mathrm{JT}}^{(i)}$,
$\delta_{\mathrm{JT}}^{(i)}$
and $\Lambda$ directly from the 
DFT potential energy surface (e.g., Fig.~\ref{fig:fig1}(d)).
The effective vibrational
energy $\hbar\omega_E$ can be found from
these parameters similarly to the case
of the single Jahn-Teller (see SI). The vibronic
splitting between the lowest levels to first
and second order are given by 
$\gamma^{(1)}$ and $\gamma^{(2)}$, respectively. 
SO effects
are included non-perturbatively and we find significant
quenching of the pure electronic SO splitting
($p_{u,g} \ll 1$), a consequence of the 
strong electron-phonon coupling induced by the 
pJT. The energy $\lambda_u + \lambda_g$ 
corresponds to the energy splitting between
the $m_s = \pm 1$ levels of the lowest $E_u$ vibronic eigenstates.
}
\label{tab:pjt}
\end{table}

\begin{figure}[t]
\includegraphics[width=\columnwidth]{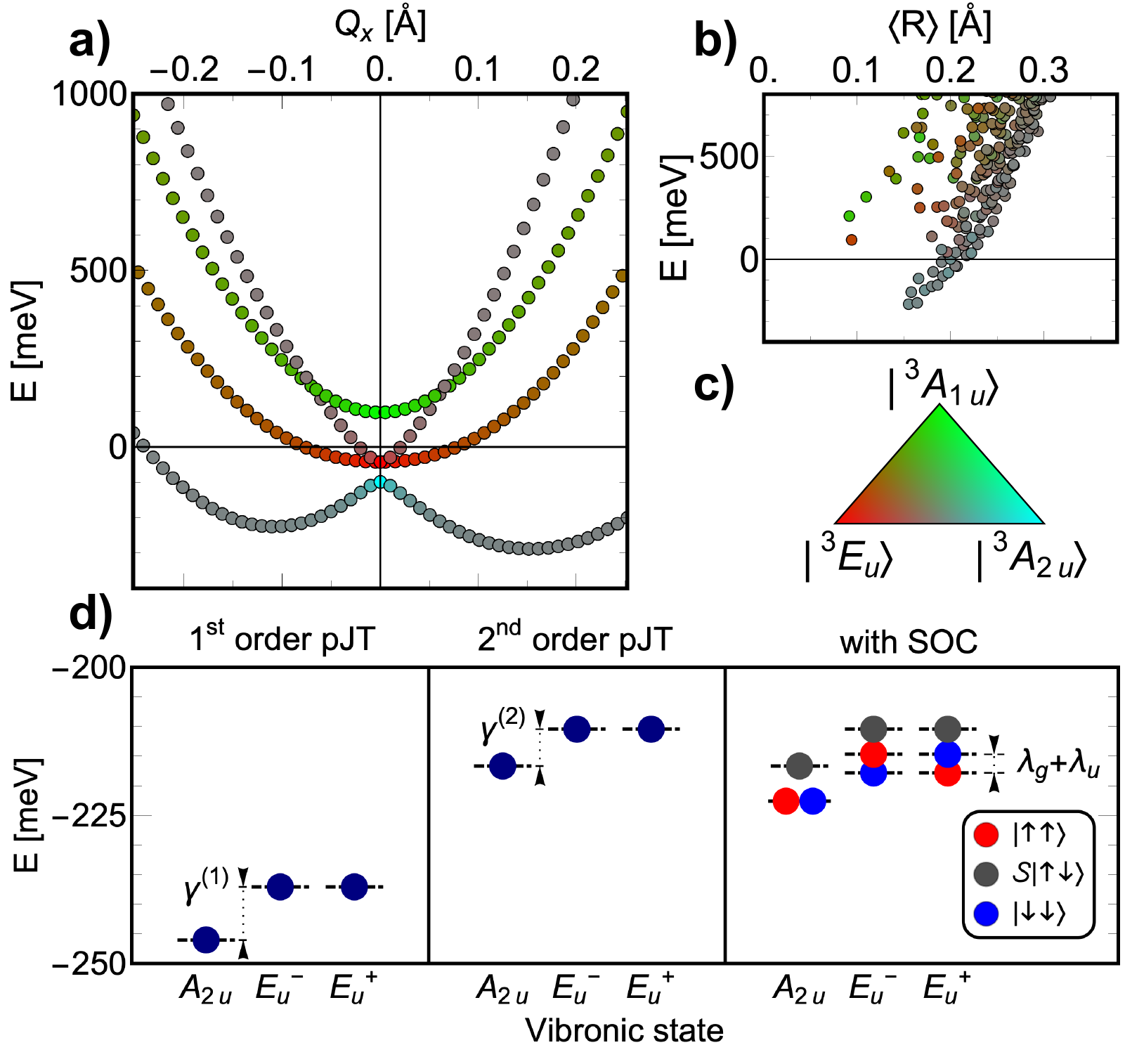}
\caption{ \textbf{(a)} 1D cut ($Q_y=0$) of the full electron-vibration coupled 
PES within the adiabatic approximation for Sn\Vn.
\textbf{(b)} 
The vibronic eigenstates found after solving the
pJT and electronic correlation Hamiltonian (Eq.~\ref{eq:pjtandcorrelation}), where the x-axis
corresponds to the expectation value of the nuclear 
position coordinate $R=\sqrt{Q_x^2+Q_y^2}$ relative to the \dddd\ minima. 
The solutions for both panels
(a) and (b) are projected onto the 
\dddd\ symmetry-adapted electronic states and the resulting composition
is represented by the color shown in \textbf{(c)}.
\textbf{(d)} The effects of 2\textsuperscript{nd} order JT and
explicit inclusion of SOC 
are detailed for the lowest-energy eigenstates of the 
system. In 
1\textsuperscript{st} and 2\textsuperscript{nd} order JT, the $A_{2u}$ state is nondegenerate
and the $E_{u}$ state is twice degenerate. The inclusion of second
order decreases the splitting $\gamma$ between these levels, while also introducing an absolute 
energy shift of around 20 meV. 
The inclusion of SOC splits the $E_u$ levels into
$E_u^+$ and $E_u^-$, each with corresponding $m_s$ sublevels. 
The splitting between the $m_s=\pm1$ levels 
is given by $\lambda_g + \lambda_u$, which is
strongly attenuated. The $m_s=0$
(labeled by $|\mathcal{S} | \uparrow \downarrow\rangle)$
levels are unaffected by SOC. 
}
\label{fig:fig2}
\end{figure}

Next we describe spin-orbit interactions in the pJT system.
In the presence of a dynamical JT
effect, expectation values of
purely electronic operators can be quenched
because of the coupled vibronic nature of the
system, as first shown by Ham~\cite{ham:1965}.
Thus it is important to analyze the
effects of SO interactions with caution,
as has already been demonstrated for the group I\Vm\
defects~\cite{thiering:gIV:2018}.
In these group I\Vn\ centers,
the SOC Hamiltonian 
can be written as a product
of the single-hole interactions~\cite{thiering:NV:2017},
since the spin-orbit coupling does not mix the 
$e_u$ and $e_g$ orbitals~\cite{heppThesis}. 
The SOC Hamiltonian is written as 
\begin{align}
    \hat{\mathrm{H}}_{\mathrm{SOC}} =&
    m_s 
    \left( 
    \frac{\lambda^0_{u}}{2} (\hat{\sigma}_y \otimes \hat{\sigma}_0)
    + \frac{\lambda^0_{g}}{2} 
    (\hat{\sigma}_0 \otimes  
    \hat{\sigma}_y) \right).
\label{eq:soc}
\end{align}
Here, we introduce SO splittings $\lambda_{u/g}^0$ 
for both the $e_u$ and $e_g$ orbitals,
which can be obtained from \emph{ab initio}
calculations. The
variable $m_s$ corresponds to eigenvalues
of $\hat{S}_z$ 
and for the $S = 1$ triplet system 
can take on values of $m_s \in [ 1, 0, -1 ]$.
While SOC in its general
$\hat{\mathbf{L}} \cdot \hat{\mathbf{S}}$ form 
(with angular momentum operator $\hat{\textbf{L}}$
and spin operator $\hat{\textbf{S}}$)
also contains transverse terms, 
these transverse terms only couple
$e_g/e_u$ orbitals to $a_{2u}$ orbitals 
which are outside the $(e_g,e_u)$ manifold of interest~\cite{heppThesis}.
This consideration allows us to effectively write 
$\hat{\mathrm{H}}_\mathrm{SOC}$ solely in terms proportional to
$\hat{L}_z\hat{S}_z$, yielding Eq.~\ref{eq:soc}. 
The $\hat{L}_z\hat{S}_z$ interactions can
couple the excited-state singlet manifold
with the $m_s=0$ excited-state triplets, however
we choose to consider only the
triplet subspace as the $(e_u^1 e_g^1)$
singlet excited states are expected to be
higher in energy due to Coulomb
repulsion~\cite{green:siv0:2017}. 
Ultimately intersystem crossing (ISC) rates between
these triplet and singlet levels will likely
depend on phonon overlaps of the full diamond + defect system,
however they require nonzero
spin-orbit coupling and thus our analysis is important
for further understanding ISC.

To capture the spin-orbit
interaction in addition to the pJT physics,
we find that including SOC 
perturbatively is insufficient, even for the Si\Vn\ system. 
Thus, we
invoke a complete spin-resolved
orbital basis including all spin
sublevels of Eq.~\ref{eq:soc}.
From this we perform direct diagonalization of the combined
spin-orbit and Jahn-Teller system (see SI), where we take all terms
in Eq.~\ref{eq:pjtandcorrelation}
to be spin-independent.
The solutions of this coupled Hamiltonian
allows us to extract both the 
absolute energy shifts of our vibronic 
eigenstates with SO effects
and the effective SO splittings
between spin sublevels non-perturbatively.

Table~\ref{tab:pjt} summarizes the results of 
our work.
In each of the defect centers studied, we find a significant 
pJT effect, with the constructive
interference yielding instabilities of over 200 meV.
We find the second order effects are also 
relatively large, with $\delta_{\mathrm{JT}}^{(1)}\sim
0.3 E_{\mathrm{JT}}^{(1)}$ for each of the defects studied.
These second order shifts are important, as they 
represent the energy barrier between the three
energy minima present in the 2D vibrational 
$(Q_x,Q_y)$ space. This energy barrier helps to determine if
the system
will prefer a static or
dynamic JT distortion,
the latter of which means
the electron and phonon degrees
of freedom cannot be decoupled
and instead a coupled vibronic solution
is required.  
Indeed, the system can be parameterized as strongly-coupled
as given by the parameter $\lambda = E_{\mathrm{JT}}/\hbar \omega_E$,
which is $> 2$ for all cases studied here. 
After calculation of the parameters in Table~\ref{tab:pjt},
we can solve for the coupled electron-vibrational
system as defined in 
Eqs.~\ref{eq:pjt}~and~\ref{eq:pjtandcorrelation}.

Figure~\ref{fig:fig2} visualizes our results for 
Sn\Vn. 
Panel (a) represents the 
adiabatic states 
along a 1D cut of the vibrational space with $Q_y = 0$. The full 
vibronic solutions to Eq.~\ref{eq:pjtandcorrelation}
are shown in panel (b), plotted as a function
of the expectation value of displacement from the high-symmetry 
\dddd\ minima.
In both cases we can project the solutions onto 
the irreducible states of the \dddd\ excited-state manifold,
with the color legend given in panel (c).
We find that the lowest energy states are comprised of roughly equal
contributions from the undistorted $|^3E_u \rangle$ and $|^3A_{2u}\rangle$ electronic states. This
is true for the quadratic coupling as well.
In Fig.~\ref{fig:fig2}(d) we specifically focus on the lowest-energy vibronic
solutions. The lowest
vibronic state has total symmetry $A_{2u}$ which is optically dark,
whereas the next eigenstate is an optically-active, 
doubly-degenerate $E_u$ level. 
In first order pJT,
the splitting $\gamma^{(1)}$ between these two states for Sn\Vn\ is 8.96 meV,
while including second-order coupling decreases splitting $\gamma^{(2)}$
to just 6.22 meV. 
Even at second order, the $^3E_u$ state remains degenerate,
however overall the eigenstates of the system shift upwards in energy by
roughly 20 meV. 

It is interesting to note
that in general including second-order terms
in the 
pJT Hamiltonian decreases
the splitting $\gamma$ between the lowest
vibronic states  
(see Table~\ref{tab:pjt}). 
This splitting was measured experimentally
for Si\Vn~\cite{green:SiV0:2019} to be
6.8 meV; here we find a larger discrepancy to experiment
in the case of quadratic coupling ($\gamma^{(2)} = 3.2$ meV)
than we do for linear coupling ($\gamma^{(1)} = 7.2$ meV). 
We emphasize, however, that an inclusion of second order electron-phonon coupling
more closely resembles the \emph{ab initio} data,
as can be seen in Fig.~\ref{fig:fig1}(d) 
due to the nonvanishing $\delta_{\mathrm{JT}}^{(i)}$. 
The origin of this disagreement
is unknown and beyond the scope
of this work.
We suggest that it may 
represent an energy-resolution limitation
in the approach
employed.
We note that inclusion of higher-order terms~\cite{viel2004} up to fourth order in
electron-phonon interactions is found to negligibly
change our results.

The coupled spin-vibronic results
are shown in the final panel of Fig.~\ref{fig:fig2}(d)
and are found after including the SOC Hamiltonian directly.
We find that the $m_s =0$ and $m_s=\pm1$ 
sublevels of the $A_{2u}$ vibronic
states are split, in the case of Sn\Vn\
by 5.9 meV.
The $m_s = \pm 1$ sublevels of the $E_u$ states also split 
(here we distinguish the $E_u$ states by 
labels $+$ and $-$). 
These $E_u^{\pm}$ states have a 
Kramers degeneracy, very much analogous to the lowest $E_g$ vibronic states
of the group I\Vm, where $|^3E_u^
+\rangle \otimes |\uparrow \uparrow \rangle$ and $
|^3E_u^-\rangle \otimes |\downarrow \downarrow \rangle$
are the degenerate, 
lowest energy $E_u$ states. These are split by an energy
of $\lambda_u + \lambda_g$ from the degenerate
$|^3E_u^
-\rangle \otimes |\uparrow \uparrow \rangle$ and $|^3E_u^+\rangle \otimes |\downarrow \downarrow \rangle$ states,
as shown in Fig.~\ref{fig:fig2}(d) for Sn\Vn. 
In the absence of JT interactions this splitting $\lambda_g+\lambda_u$
would be over 100 meV, however here it is only $\sim3$ meV, a direct consequence
of the strong electron-phonon coupling present in the pJT system. 
Additional interactions
such as effects of strain and spin-spin coupling could 
split and shift these levels further.

For all cases, the reduction factors denoted by $p_{u/g}$ are 
smaller than 0.05, indicative of a very strong quenching of the
SO interaction, even more so than the group I\Vm\ color centers.
This can be attributed in part to the scaling of the 
Jahn-Teller instability vs. the spin-orbit splitting in the two-hole case.
While to first order the JT energy scales as the square of the coupling
(i.e., $\sim (F_u + F_g)^2$), the SO splitting scales linearly (i.e.,
$\lambda_g + \lambda_u$). Such a scaling and the resulting JT energies 
intuitively explains the significant SO quenching we find in this work. 
We note that shifts in the absolute energies of the $E_u$
states are found to be most significant in the case of Pb\Vn,
where we find a redshift in the predicted ZPL of roughly 0.05 eV. All lighter
defects have much weaker absolute energy shifts due to their reduced 
SO coupling constants.

In conclusion, we present first principles calculations of group IV neutral
artificial atoms in diamond, where
we capture the product Jahn-Teller effect to second order in
electron-phonon coupling and non-perturbatively describe the effects
of spin-orbit interactions. 
Our results find significant reduction in the spin-orbit splitting 
due to the strong pJT. However, we also find that the spin-orbit
interactions would split the lowest optically-active states into
$m_s$-resolved levels split by up to a few meV in the heavier 
candidates. 
These results provide qualitatively new insight into the physics
of artificial atom qubits in diamond and are of quantitative importance
in experimental identification and manipulation of these centers in quantum information science.

\section*{Acknowledgments}
The authors thank Dr. Tom\'a\v{s} Neuman and Prof. Marko Lon\v{c}ar, at Harvard University, for helpful discussions.

This work was supported by the DOE `Photonics at Thermodynamic Limits' Energy Frontier Research Center under grant number DE-SC0019140.
D. E. and P.N. are partially supported by the Army Research Office MURI (Ab-Initio Solid-State Quantum Materials)
grant number W911NF-18-1-0431 and by the STC Center for Integrated Quantum Materials (CIQM) under NSF grant number DMR-1231319. This research used resources of the National Energy Research
Scientific Computing Center, a DOE Office of Science User Facility
supported by the Office of Science of the U.S. Department of Energy
under Contract No. DE-AC02-05CH11231. Additional calculations were performed using resources from the Department of Defense High Performance Computing
Modernization program as well as resources at the Research Computing
Group at Harvard University.
J. F. acknowledges partial financial support from the Deutsche Forschungsgemeinschaft (DFG) under contract No. FL 997/1-1. The Flatiron Institute is a division of the Simons Foundation. P.N. is a Moore Inventor Fellow.

\end{document}